**Studying and Enhancing Talking Condition Recognition in Stressful and Emotional Talking Environments Based on HMMs, CHMM2s and SPHMMs**

Ismail Shahin

Electrical and Computer Engineering Department

University of Sharjah

P. O. Box  27272

Sharjah, United Arab Emirates

Tel: (971) 6 5050967

Fax: (971) 6 5050877

E-mail: ismail@sharjah.ac.ae

**Abstract**

The work of this research is devoted to studying and enhancing talking condition recognition in stressful and emotional talking environments (completely two separate environments) based on three different and separate classifiers. The three classifiers are: Hidden Markov Models (HMMs), Second-Order Circular Hidden Markov Models (CHMM2s) and Suprasegmental Hidden Markov Models (SPHMMs). The stressful talking environments that have been used in this work are composed of neutral, shouted, slow, loud, soft and fast talking conditions, while the emotional talking environments are made up of neutral, angry, sad, happy, disgust and fear emotions. The achieved results in the current work show that SPHMMs lead each of HMMs and CHMM2s in improving talking condition recognition in stressful and emotional talking environments. The results also demonstrate that talking condition recognition in stressful talking environments outperforms that in emotional talking environments by 2.7%, 1.8% and 3.3% based on HMMs, CHMM2s and SPHMMs, respectively. Based on subjective assessment by human judges, the recognition performance of stressful talking conditions leads that of emotional ones by 5.2%.

## 1. Introduction

Talking condition/emotion recognition by machine (computer) is the task of recognizing the unknown talking condition/emotion based on the information embedded in the speech signal. Talking condition/emotion recognition is divided into talking condition/emotion identification and talking condition/emotion verification (authentication). In talking condition/emotion identification, the unknown talking condition/emotion is identified as the talking condition/emotion whose model best matches the input speech signal. In talking condition/emotion verification, the aim is to decide whether a given talking condition/emotion corresponds to a particular known talking condition/emotion or to some other unknown talking conditions/emotions. Based on the text to be spoken, talking condition/emotion recognition methods can be classified into text-dependent or text-independent. Text-dependent (fixed-text) talking condition/emotion recognition requires to generate speech of the same text in both training and testing under a talking condition/emotion; on the other hand, text-independent (free-text) does not depend on the text being spoken under a talking condition/emotion.

## 2. Motivation and Literature Review

Speech communication is one of the most important channels between users and machines (computers) and it can be used to recognize the talking condition/emotional state of a speaker. The talking condition/emotional state of a speaker can be recognized through his/her: facial expression, speech signal uttered by the speaker, gesture, heart rate, temperature and blood pressure. One important research challenge in the last decade has thus been automatically recognizing the talking condition/emotional status of a speaker using speech. Talking condition recognition in each of stressful and emotional talking environments is a vital research field for human-computer interaction [1]. A major motivation comes from the demand to develop human-computer interface that is more adaptive and reactive to a user's talking condition/emotion. The major task of intelligent human-machine interaction is to enable the computer with the human-computer interaction capability so that the computer can recognize the talking condition/emotion of the user for a wide range of different applications.

Stressful talking environments are defined as the talking environments where speakers generate their speech under the impact of stressful circumstances such as shouted circumstance. Some factors that introduce stress into the speech production process include noisy background, emergency conditions such as that in aircraft pilot communications, high workload stress, physical environmental factors, multitasking and physical G-force movement such as fighter cockpit pilot [2]. Talking condition recognition in stressful talking environments has many applications. Such applications include emergency telephone message sorting, telephone banking, hospitals which include computerized stress classification and assessment techniques and military voice communication and control applications.

There are many studies in the field of stressful talking condition recognition [2], [3], [4], [5]. Some talking conditions are designed to simulate speech under real stressful talking conditions. The authors of *Refs.*[2] and [5] recorded and used Speech Under Simulated and Actual Stress (SUSAS) database in which eight talking conditions are used to simulate speech produced under real stressful talking conditions [2], [5]. The eight talking conditions are neutral, loud, soft, angry, fast, slow, clear and question. The author of *Ref.* [3] studied talking condition identification using circular hidden Markov models (CHMMs). He used neutral, shouted, loud, slow and fast talking conditions [3]. The author of *Ref.* [4] studied talker-stress-induced intraword variability and an algorithm that compensates for the systematic changes observed based on hidden Markov models (HMMs) trained by speech tokens in various talking conditions. He used six talking conditions to simulate speech under real stressful talking conditions. The talking conditions are neutral, fast, loud, Lombard, soft and shouted [4].

Emotional talking environments can be defined as the talking environments where speakers produce their speech under the influence of emotional situations such as anger, happiness and sadness. Emotion recognition can be used in many applications. Such applications appear in telecommunications, human robotic interfaces, smart call centers and intelligent spoken tutoring systems. In telecommunications, emotion recognition can be used to assess a caller's emotional state for telephone response services. Emotion recognition can also be used in human robotic interfaces, where robots can be taught to intermingle with humans and recognize human emotions. More applications of emotion recognition from speech appear in smart call centers where emotion recognition can help to spot possible problems emerging from an unsatisfactory interaction. In intelligent spoken tutoring systems, emotion recognition can be employed to detect and adapt to students' emotions when students got bored during the tutoring session.

There are many studies in the field of emotion recognition. The authors of *Ref.* [6] shed the light on recognizing emotions from spoken language [6]. They used a combination of three sources of information for emotion recognition. The three sources are acoustic, lexical and discourse. The authors of *Ref.* [7] aimed in one of their work to enhance the automatic emotional speech classification methods using ensemble or multi-classifier system (MCS) approaches. They also aimed to examine the differences in perceiving emotion in human speech that is derived from different methods of acquisition [7]. The authors of *Ref.* [8] proposed in one of their studies a text-independent method of emotion classification of speech based on HMMs [8]. The authors of *Ref.* [9] proposed a new feature vector that helps in enhancing the classification performance of emotional/stressful states of humans. The elements of such a feature vector are achieved from a feature subset selection method based on genetic algorithm [9].

In literature, different techniques, algorithms and models have been employed to classify the stressful/emotional state of a speaker through speech. HMMs have been used by the authors of *Refs.* [2], [3], [8]. Neural Networks (NNs) have been applied by the authors of *Refs.* [10], [11]. Genetic Algorithms (GAs) have been exploited by the authors of *Ref.* [9]. Support Vector Machines (SVMs) have been implemented by the authors of *Refs.* [12], [13]. Our main contribution in this work is focused on studying and enhancing text-independent and speaker-independent talking condition identification in stressful and emotional talking environments (completely two separate environments) based on three separate and different classifiers. The three classifiers are: HMMs (well-known models), Second-Order Circular Hidden Markov Models (CHMM2s) and Suprasegmental Hidden Markov Models (SPHMMs). The last two models have been developed, used and evaluated by the author of *Refs.* [14], [15]. The stressful talking environments used in the current work consist of six stressful talking conditions. The talking conditions are neutral, shouted, slow, loud, soft and fast. On the other hand, the emotional talking environments used in the present work are composed of six emotions. These emotions are neutral, angry, sad, happy, disgust and fear. In addition, one of our main contributions in this work is to discriminate between stressful talking condition environment and emotional talking condition environment. This work is a multidisciplinary area involving two fields: speech signal processing and human-machine interaction.

The remainder of this paper is organized as follows. Section 3 covers the fundamentals of CHMM2s. The basics of SPHMMs are given in Section 4. The speech databases used in this work and the extraction of features are explained in Section 5. Section 6 discusses the algorithm of talking condition/emotion identification based on HMMs, CHMM2s and SPHMMs. Section 7 discusses the results obtained in this work and the experiments. Finally, Section 8 concludes this work with some remarks.

### 3. Fundamentals of Second-Order Circular Hidden Markov Models

In one of his studies, the author of *Ref.* [14] proposed, implemented and tested CHMM2s to improve speaker identification performance in shouted talking environments [14]. CHMM2s have proven to be superior models over each of: first-order left-to-right hidden Markov models (LTRHMM1s), second-order left-to-right hidden Markov models (LTRHMM2s) and first-order circular hidden Markov models (CHMM1s). The reason of superiority is that CHMM2s possess the characteristics of both CHMMs and HMM2s [14].

The initial elements of the parameters in the training phase of CHMM2s are chosen to be [14],

$$v_k(i) = \frac{1}{N} \qquad N \geq i, k \geq 1 \qquad (1)$$

where $v_k(i)$ is the initial element of the probability of an initial state distribution.

$$\alpha_1(i,k) = v_k(i) b_{ki}(O_1) \qquad N \geq i, k \geq 1 \qquad (2)$$

where $\alpha_1(i,k)$ is the initial element of the forward probability of producing the observation vector $O_1$.

$$a_{ijk}^1 = \begin{cases} \dfrac{1}{3} & i = 1, j, k = 1, 2, ..., N \\ \\ \dfrac{1}{3} & N-1 \geq i \geq 2, i+1 \geq j \geq i-1, N \geq k \geq 1 \\ \\ \dfrac{1}{3} & i = N, j, k = 1, 2, ..., N \\ \\ 0 & \text{otherwise} \end{cases} \qquad (3)$$

where $a^1_{ijk}$ is the initial element of $a_{ijk}$.

$$b^1_{ijk} = \frac{1}{M} \qquad N \geq j, k \geq 1, M \geq i \geq 1 \qquad (4)$$

where $b^1_{ijk}$ is the initial element of the observation symbol probability and $M$ is the number of observation symbols.

$$\beta_T(j,k) = \frac{1}{N} \qquad N \geq j, k \geq 1 \qquad (5)$$

where $\beta_T(j,k)$ is the initial element of the backward probability of generating the observation vector $O_T$.

$$P(O|\Phi) = \sum_{k=1}^{N} \sum_{i=1}^{N} \alpha_T(i,k) \qquad (6)$$

where $P(O|\Phi)$ is the probability of the observation vector $O$ given the CHMM2s model $\Phi$. More details about the second-order circular hidden Markov models can be found in *Ref.* [14].

## 4. Basics of Suprasegmental Hidden Markov Models

SPHMMs have been developed, used and evaluated by the author of *Refs.* [15], [16], [17] in the fields of speaker recognition [15], [16] and emotion recognition [17]. SPHMMs have the ability to encapsulate several states of HMMs into what is labeled a suprasegmental state. Suprasegmental state deals with the observation sequence through a larger window. This suprasegmental state permits observations at appropriate rates for the situation of modeling. Prosodic information, for example, can not be detected at a rate that is used for acoustic modeling. The main acoustic parameters that express prosody are fundamental frequency, intensity and duration of speech signals [18]. The prosodic features of a unit of speech are characterized as suprasegmental features because they have influence on all the segments of the unit of speech. Therefore, prosodic events at the levels of phone, syllable, word and utterance are represented using suprasegmental states; on the other hand, acoustic events are represented using conventional hidden Markov states.

Within HMMs, prosodic and acoustic information can be combined as given by the following formula [19],

$$log\ P(\lambda^v, \Psi^v | O) = (1-\alpha).\ log\ P(\lambda^v | O) + \alpha.\ log\ P(\Psi^v | O) \quad (7)$$

where $\alpha$ is a weighting factor. When

$$\begin{cases} 0.5 > \alpha > 0 & \text{biased towards acoustic model} \\ 1 > \alpha > 0.5 & \text{biased towards prosodic model} \\ 0 & \text{biased completely towards acoustic model and} \\ & \text{no effect of prosodic model} \\ 0.5 & \text{no biasing towards any model} \\ 1 & \text{biased completely towards prosodic model and} \\ & \text{no impact of acoustic model} \end{cases} \quad (8)$$

$\lambda^v$: is the acoustic model of the $v$th talking condition/emotion.

$\Psi^v$: is the suprasegmental model of the $v$th talking condition/emotion.

$O$: is the observation vector or sequence of an utterance. The reader can obtain more information about suprasegmental hidden Markov models from *Ref*. [15].

## 5. Speech Databases and Extraction of Features

### 5.1 Speech Databases

Each of the stressful and emotional speech databases was collected from 30 (15 male and 15 female) non-professional (therefore, our speech database is closer to the real-life data than to the acted data) healthy adult Native American speakers. Each speaker uttered 8 sentences where each sentence was uttered 9 times under each one of the 6 stressful talking conditions (neutral, shouted, slow, loud, soft and fast) and each one of the 6 emotions (neutral, angry, sad, happy, disgust and fear).

The total number of utterances per talking environment was 12960. These sentences are:

1) *He works five days a week.*
2) *The sun is shining.*
3) *The weather is fair.*
4) *The students study hard.*
5) *Assistant professors are looking for promotion.*
6) *University of Sharjah.*
7) *Electrical and Computer Engineering Department.*
8) *He has two sons and two daughters.*

The two speech databases in this work were captured separately by a speech acquisition board using a 16-bit linear coding A/D converter and sampled at a sampling rate of 16 kHz. These databases were 16-bit per sample linear data.

## 5.2 Extraction of Features

The features that represent the phonetic content of speech signals in the two databases of this work are called the Mel-Frequency Cepstral Coefficients (static MFCCs) and delta Mel-Frequency Cepstral Coefficients (delta MFCCs). These coefficients have been extensively used in many studies in the areas of speech recognition [5], [20], [21], speaker recognition [22], [23], and emotion recognition [6], [13], [24]. These coefficients outperform other coefficients in the three areas. MFCC feature analysis was used to form the observation vectors for each of HMMs, CHMM2s and SPHMMs in the stressful and emotional talking environments.

The number of conventional states in each of HMMs and CHMM2s was 9, while the number of suprasegmental states in SPHMMs was 3 (each suprasegmental state was composed of 3 conventional states). The number of mixture components, *M*, was 10 per state, with a continuous mixture observation density was selected for each model.

## 6. Talking Condition/Emotion Identification Algorithm Based on HMMs, CHMM2s and SPHMMs

### 6.1 Talking Condition/Emotion Identification Algorithm Based on HMMs

HMMs have become popular in the fields of speech recognition and speaker recognition in the last few decades [25], [26]. Recently, HMMs have been used in the field of talking condition/emotion recognition [3], [7], [8], [27], [28]. Left-to-Right Hidden Markov Models (LTRHMMs) have been adopted in this work.

The probability of the observation vector *O* given the HMM stress/emotion model $\lambda$, can be calculated as [25],

$$P(O|\lambda) = \sum_{i=1}^{N} \alpha_T(i) \tag{9}$$

where $\alpha_T(i)$ is the terminal forward variable that can be determined by the forward algorithm and *N* is the number of states of the model. The details of the training and re-estimation algorithms can be found in many *Refs.* [25], [26].

In the training session in each of the stressful and emotional talking environments (completely two separate training sessions), one reference model per talking condition/emotion was derived using 20 out of the 30 speakers uttering the first 4 sentences of the 8 sentences where each sentence was uttered 9 times. Therefore, each talking condition/emotion was represented by one reference talking condition/emotion model. The number of utterances used in this session to derive each HMM talking condition/emotion model was 720 per talking environment (20 speakers x 4 sentences x 9 utterances/sentence).

In the test (identification) session in each of the stressful and emotional talking environments (completely two separate test sessions), each one of the 10 remaining speakers used the second 4 sentences with a repetition of 9 utterances/sentence under each talking condition/emotion (text-independent and speaker-independent). The number of utterances used in this session was 2160 per talking environment (10 speakers x 4 sentences x 9 utterances/sentence x 6 talking conditions/emotions). The probability of generating every utterance was computed based on HMMs (there were 6 probabilities per talking environment), the model with the highest probability was chosen as the output of talking condition/emotion identification as given in the following formula,

$$V^* = \arg\max_{6 \geq v \geq 1} \left\{ P\left(O \big| \lambda^v \right) \right\} \quad (10)$$

where,

$V^*$: is the index of the identified talking condition/emotion.

$O$: is the observation vector or sequence of the unknown talking condition/emotion.

$P(O|\lambda^v)$: is the probability of the observation sequence $O$ given the $v$th HMM talking condition/emotion model $\lambda^v$.

## 6.2 Talking Condition/Emotion Identification Algorithm Based on CHMM2s

In the training session in each of the stressful and emotional talking environments, one reference model per talking condition/emotion was built using the 20 speakers generating the first 4 sentences where each sentence was produced 9 times. The number of utterances used in this session to construct each CHMM2 talking condition/emotion model was 720.

In the test session in each of the stressful and emotional talking environments, each one of the 10 remaining speakers used the second four sentences with a repetition of 9 times under each talking condition/emotion (text-independent and speaker-independent). The number of utterances used in this session was 2160 per talking environment. The probability of generating every utterance was computed based on CHMM2s, the model with the highest probability was chosen as the output of talking condition/emotion identification in each of the stressful and emotional talking environments as given in the following formula,

$$V^* = \arg\max_{6 \geq v \geq 1} \left\{ P\left(O \big| \Phi^v \right) \right\} \quad (11)$$

where $P(O|\Phi^v)$ is the probability of the observation sequence $O$ given the $v$th CHMM2 talking condition/emotion model $\Phi^v$.

## 6.3 Talking Condition/Emotion Identification Algorithm Based on SPHMMs

Since phonemes follow strictly the left to right sequence, most of studies in speech and speaker recognition areas used the Left-to-Right HMMs (LTRHMMs). In this work, Left-to-Right Suprasegmental Hidden Markov Models (LTRSPHHMs) were derived from LTRHMMs. Fig. 1 shows an example of a basic structure of LTRSPHMMs that was derived from LTRHMMs. In this figure, $q_1, q_2, ..., q_6$ are conventional hidden Markov states, $p_1$ is a suprasegmental state that consists of the states: $q_1$, $q_2$ and $q_3$, $p_2$ is a suprasegmental state that is made up of the states: $q_4$, $q_5$ and $q_6$, $p_3$ is a suprasegmental state that is composed of $p_1$ and $p_2$, $a_{ij}$ is the transition probability between the *i*th conventional hidden Markov state and the *j*th conventional hidden Markov state and $b_{ij}$ is the transition probability between the *i*th suprasegmental state and the *j*th suprasegmental state.

The training session of SPHMMs in each of the stressful and emotional talking environments was similar to the training session of the conventional HMMs. In the training session of SPHMMs, suprasegmental models were trained on top of acoustic models. In each of the stressful and emotional talking environments, one reference model per talking condition/emotion was obtained using the 20 speakers speaking the first 4 sentences with a repetition of 9 times per sentence. The number of utterances used in this stage to derive each SPHMM talking condition/emotion model was 720.

In the test session in each talking environment, each one of the 10 remaining speakers used the second 4 sentences with a repetition of 9 times under each talking condition/emotion (text-independent and speaker-independent). The total number of utterances used in this stage was 2160 per talking environment. The probability of generating every utterance was computed based on SPHMMs as given in the following formula,

$$V^* = \arg\max_{6 \geq v \geq 1} \left\{ P\left(O \mid \lambda^v, \Psi^v \right) \right\} \quad (12)$$

where $P\left(O \mid \lambda^v, \Psi^v\right)$ is the probability of the observation sequence *O* given the *v*th SPHMM talking condition/emotion model *($\lambda^v$, $\Psi^v$)*.

## 7. Results and Discussion

In the current work, talking condition identification has been studied and improved in stressful and emotional talking environments based on each of HMMs, CHMM2s and SPHMMs. To the best of our knowledge, this work is the first attempt to employ CHMM2s and SPHMMs to study and enhance talking condition identification performance in stressful and emotional talking environments. In SPHMMs, the weighting factor ($\alpha$) is chosen to be equal to 0.5 to avoid biasing towards either acoustic or prosodic model.

Table 1 summarizes talking condition identification performance in stressful talking environments based on each of HMMs, CHMM2s and SPHMMs using the collected database. This table shows that the average talking condition identification performance in such talking environments based on HMMs, CHMM2s and SPHMMs is 63.8%, 68.1% and 72.0%, respectively. The table shows that SPHMMs significantly enhance the identification performance compared to HMMs and CHMM2s by 12.9% and 5.7%, respectively. Fig. 2 shows relative improvement per each stressful talking condition of using SPHMMs over each of HMMs and CHMM2s. It is apparent from this figure that the highest relative improvement happens under the slow talking condition (19.2%), while the least relative enhancement occurs under the neutral talking condition (1.6%).

Table 2, Table 3 and Table 4 show confusion matrices that represent the percentage of confusion of a test stressful talking condition with the other stressful talking conditions based on HMMs, CHMM2s and SPHMMs, respectively. Taking Table 2 as an example, this table shows the following:

a) The most easily recognizable stressful talking condition based on HMMs is neutral (92.0%). Therefore, the highest talking condition identification performance in stressful talking environments based on such models is neutral.

b) The least easily recognizable stressful talking condition based on HMMs is shouted (50.5%). So, the least talking condition identification performance in stressful talking environments based on these models is shouted.

c) The third column ('Slow' column), for example, shows that 7% of the utterances that were portrayed in a slow talking condition were evaluated as produced in a shouted talking condition, 4% of the utterances that were uttered in a slow talking condition were identified as generated in a loud talking condition. This column shows that slow talking condition has the highest confusion percentage with soft talking condition (20%). Therefore, slow talking condition is highly confusable with soft talking condition. This column also shows that slow talking condition has the least confusion percentage with fast talking condition (3%). Therefore, slow talking condition is rarely confusable with fast talking condition.

Using the collected emotional database, emotion identification performance based on each of HMMs, CHMM2s and SPHMMs is demonstrated in Table 5. This table yields average emotion identification performance of 62.1%, 66.9% and 69.7% based on HMMs, CHMM2s and SPHMMs, respectively. This table shows that SPHMMs significantly improve emotion identification performance compared to HMMs and CHMM2s by 12.2% and 4.2%, respectively. Fig. 3 illustrates relative improvement per each emotion of using SPHMMs over each of HMMs and CHMM2s in emotional talking environments. It is evident from this figure that the highest relative improvement takes place under the angry emotion (25.6%), while the least relative enhancement occurs under the neutral emotion (1.1%). Confusion matrices that represent percentage of confusion of a test emotion with the other emotions based on HMMs, CHMM2s and SPHMMs are given, respectively, in Table 6, Table 7 and Table 8.

Both of stressful talking condition identification and emotion identification performances are low based on HMMs. It is evident that HMMs are inconvenient and not powerful enough as classifiers for each of stressful talking condition identification and emotion identification. The author of *Ref.* [3] achieved in one of his studies 60.8% as an average talking condition identification performance based on circular HMMs [3]. In another study by the same author, he obtained an average talking condition identification performance of 54.8% based on second-order HMMs [29].

Based on CHMM2s, the performance of each of stressful talking condition identification and emotion identification has been greatly enhanced compared to that based on HMMs. The reason is that CHMM2s possess the characteristics of both CHMMs and HMM2s [14].

Comparing SPHMMs with each of HMMs and CHMM2s, it is apparent that SPHMMs outperform each of HMMs and CHMM2s for each of stressful talking condition identification and emotion identification. This may be attributed to the reason that SPHMMs have the ability to integrate observations from talking condition/emotional modality because such models allow for observations at an appropriate rate for stressful and emotional speech signals.

The results reported in Table 1 and Table 5 show evidently that there is a significant difference between stressful talking condition identification performance and emotional talking condition identification performance. The average talking condition identification performance in stressful talking environments based on HMMs, CHMM2s and SPHMMs is 63.8%, 68.1% and 72.0%, respectively; on the other hand, the average emotion identification performance in emotional talking environments based on HMMs, CHMM2s and SPHMMs is 62.1%, 66.9% and 69.7%, respectively. Therefore, the average stressful talking condition identification performance based on HMMs, CHMM2s and SPHMMs is better than the average emotion identification performance by 2.7%, 1.8% and 3.3%, respectively. This may be attributed to a number of reasons:

1. HMMs are more powerful and more efficient in stressful talking environments than in emotional talking environments. HMMs do not represent the changing statistical characteristics that exist in the actual observations of speech signals in emotional talking environmens as they do in stressful talking environments. This is because it is commonly believed that emotion speech is effectively represented by prosodic features, while stress speech is efficiently represented by acoustic features. Therefore, HMMs can not represent prosodic features effectively.

2. Emotions in emotional talking environments are not simple phenomena, and many factors contribute to them. A complete definition of emotions must take into account the experience feeling of emotions, the processes that occur in the brain and nervous system and the observable expressive patterns of emotions [30].

In this work, our results of stressful talking condition identification performance and emotion identification performance are better than those reported in previous studies:

1) The authors of *Ref.* [13] obtained 70.1% as an average talking condition identification performance for 4-class talking condition classification based on Gaussian SVM using SUSAS database. They also achieved, using AIBO database, 42.3% as an average emotion identification performance for 5-class emotion identification [13].

2) The author of *Ref.* [12] reported an average emotion identification performance of 55.6% using 5 emotions based on an unsupervised series experiment [12].

3) The authors of *Ref.* [9] attained an average 4-stressful talking condition identification performance of 44.6% of text-independent multistyle classification using MFCCs, while they obtained an average 4-stressful talking condition identification performance of 66.0% of text-independent multistyle classification using 16-GA feature [9].

Three more experiments have been separately conducted to evaluate the achieved results. The three experiments are:

i) Experiment 1: Talking condition recognition in stressful and emotional talking environments has been evaluated on well-known speech databases called SUSAS and Emotional Prosody Speech and Transcripts databases, respectively. These databases were produced by the Linguistic Data Consortium (LDC).

SUSAS database was designed originally for speech recognition under neutral and stressful talking conditions [31]. The database is divided into five domains, encompassing a wide range of stresses and emotions. A total of 32 speakers (13 female and 19 male), with ages ranging from 22 to 76 were employed to generate more than 16,000 utterances. The five stress domains include: i) talking styles (slow, fast, soft, loud, angry, clear and question), ii) single tracking task or speech produced in noise (Lombard effect), iii) dual tracking computer response task, iv) actual subject motion-fear tasks (G-force, Lombard effect, noise and fear), v) psychiatric analysis data (speech under depression, fear and anxiety). The first SUSAS database domain involves speech under various talking conditions. This portion of SUSAS contains utterances from 9 male speakers under 8 talking conditions (neutral, slow, fast, soft, loud, question, clear and angry). In this work, only 8 different words (4 words were used for training and the rest were used for testing) uttered by 9 male speakers (5 speakers were used for training and 4 for testing) talking in 6 talking conditions were used. The 6 talking conditions are neutral, angry, slow, loud, soft and fast.

Emotional Prosody data corpus is composed of 8 professional speakers (3 actors and 5 actresses) uttering a series of semantically neutral utterances composed of dates and numbers spoken in 15 different emotions including the neutral state. These emotions are neutral, hot anger, cold anger, panic, anxiety, despair, sadness, elation, happiness, interest, boredom, shame, pride, disgust and contempt [32]. In this work, only 8 different sentences (4 sentences were used for training and the remaining were used for testing) uttered by 8 speakers (5 speakers were used for training and the rest were used for testing) talking in 6 emotions were used. These emotions are neutral, hot anger, sadness, happiness, disgust and panic.

Table 9 demonstrates talking condition identification performance using SUSAS database based on each of HMMs, CHMM2s and SPHMMs. This table yields average talking condition identification performance 64.8%, 69.0% and 72.8% based on HMMs, CHMM2s and SPHMMs, respectively. It is evident from Table 1 (using our collected stressful database) and Table 9 (using SUSAS database) that the stressful talking condition identification performances are very close.

Table 10 shows emotion identification performance using Emotional Prosody database based on each of HMMs, CHMM2s and SPHMMs. This table gives average emotion identification performance 63.6%, 67.9% and 70.4% based on HMMs, CHMM2s and SPHMMs, respectively. It is apparent from Table 5 (using our collected emotion database) and Table 10 (using Emotional Prosody database) that the emotion identification performances are very similar.

ii) Experiment 2: Talking condition recognition in stressful and emotional talking environments has been assessed for distinct values of the weighting factor ($\alpha$). Fig. 4 and Fig. 5 demonstrate average talking condition identification performance for different values of $\alpha$ (0.0, 0.1, …, 0.9, 1.0) in the stressful and emotional talking environments, respectively. These two figures show that as the value of the weighting factor increases, the average talking condition identification performance (excluding the neutral talking condition) enhances significantly. Therefore, it is evident that SPHMMs have more impact on talking condition identification performance than HMMs.

iii) Experiment 3: An informal subjective evaluation for each of stressful talking condition identification and emotion identification (completely two separate evaluations) using the collected speech database has been conducted by 10 listeners (human judges). These listeners are non-professional healthy adult Native American speakers. A total of 720 utterances per talking environment (30 speakers x 6 talking conditions/emotions x 4 sentences only) have been used in this evaluation. During each evaluation, the 10 listeners are asked to identify the unknown talking condition/emotion. The average performance of stressful talking condition identification and emotion identification is 68.9% and 65.5%, respectively. The two averages are close to the achieved averages in the current work. It is evident from the two averages that the performance of stressful talking condition identification leads that of emotion identification.

## 8. Concluding Remarks

In this work, the focus is on studying and enhancing stressful and emotional talking condition identification performance based on each of HMMs, CHMM2s and SPHMMs. The current work is a multidisciplinary area that includes two fields of study: speech signal processing and human-machine interaction. Some conclusions can be drawn in this work. Firstly, SPHMMs are superior models over each of HMMs and CHMM2s for stressful talking condition identification and emotion identification. Secondly, the results of this work show evidently that stressful talking condition identification performs better than emotion identification based on each of HMMs, CHMM2s and SPHHMs. Finally, our results are limited. Therefore, a new model or approach is required to discriminate between stressful talking condition environment and emotional talking condition environment.

# References

[1] R. W. Picard, "Affective computing," MIT Media Lab Perceptual Computing Section Tech. Rep., No. 321, 1995.

[2] S. E. Bou-Ghazale and J. H. L. Hansen, "A comparative study of traditional and newly proposed features for recognition of speech under stress," IEEE Transaction on Speech & Audio Processing, Vol. 8, No. 4, July 2000, pp. 429-442.

[3] I. Shahin, "Talking condition identification using circular hidden Markov models," 2$^{nd}$ International Conference on Information & Communication Technologies: from Theory to Applications (ICTTA'06, IEEE Section France), Damascus, Syria, April 2006.

[4] Y. Chen, "Cepstral domain talker stress compensation for robust speech recognition," IEEE Transaction on ASSP, Vol. 36, No. 4, April 1988, pp. 433-439.

[5] G. Zhou, J. H. L. Hansen and J. F. Kaiser, "Nonlinear feature based classification of speech under stress," IEEE Transaction on Speech & Audio Processing, Vol. 9, No. 3, March 2001, pp. 201-216.

[6] C. M. Lee and S. S. Narayanan, "Towards detecting emotions in spoken dialogs," IEEE Transaction on Speech and Audio Processing, Vol. 13, No. 2, March 2005, pp. 293-303.

[7] D. Morrison, R. Wang, and L. C. De Silva, "Ensemble methods for spoken emotion recognition in call-centres," Speech Communication, Vol. 49, issue 2, February 2007, pp. 98-112.

[8] T. L. Nwe, S. W. Foo, L. C. De Silva, "Speech emotion recognition using hidden Markov models," Speech Communication, Vol. 41, issue 4, November 2003, pp. 603-623.

[9] S. Casale, A. Russo, S. Serrano, " Multistyle classification of speech under stress using feature subset selection based on genetic algorithms," Speech Communication, Vol. 49, issues 10-11, October-November 2007, pp. 801-810.

[10] J. H. L. Hansen and B. Womack, "Feature analysis and neural network-based classification of speech under stress," IEEE Transactions on Speech and Audio Processing, Vol. 4, No. 4, 1996, pp. 307-313.

[11] C. H. Park and K.B. Sim, "Emotion recognition and acoustic analysis from speech signal," Proceedings of the International Joint Conference on Neural Networks, Vol. 4, July 20-24, 2003, Portland, Oregon, USA, pp. 2594-2598.

[12] P.-Y. Oudeyer, "The production and recognition of emotions in speech: features and algorithms," International Journal of Human-Computer Studies, Vol. 59, 2003, pp. 157-183.

[13] O. W. Kwon, K. Chan, J. Hao, T. W. Lee, "Emotion recognition by speech signals," 8$^{th}$ European Conference on Speech Communication and Technology 2003, Geneva, Switzerland, September 2003, pp. 125-128.

[14] I. Shahin, "Enhancing speaker identification performance under the shouted talking condition using second-order circular hidden Markov models," Speech Communication, Vol. 48, issue 8, August 2006, pp. 1047-1055.

[15] I. Shahin, "Speaker identification in the shouted environment using suprasegmental hidden Markov models," Signal Processing Journal, Vol. 88, issue 11, November 2008, pp. 2700-2708.

[16] I. Shahin, "Speaker identification in emotional environments," Iranian Journal of Electrical and Computer Engineering, Vol. 8, No. 1, Winter-Spring 2009, pp. 41-46.

[17] I. Shahin, "Speaking style authentication using suprasegmental hidden Markov models," University of Sharjah Journal of Pure and Applied Sciences, Vol. 5, No. 2, June 2008, pp. 41-65.

[18] J. Adell, A. Benafonte, and D. Escudero, "Analysis of prosodic features: towards modeling of emotional and pragmatic attributes of speech," XXI Congreso de la Sociedad Española para el Procesamiento del Lenguaje Natural, SEPLN, Granada, Spain, September 2005.

[19] T. S. Polzin and A. H. Waibel, "Detecting emotions in Speech," Cooperative Multimodal Communication, Second International Conference 1998, CMC 1998.

[20] S. Davis and P. Mermelstein, "Comparison of parametric representations for monosyllabic word recognition in continuously spoken sentences," IEEE Transactions on Acoustics, Speech and Signal Processing, Vol. 28, issue 4, August 1980, pp. 357-366.

[21] V. Pitsikalis and P. Maragos, "Analysis and classification of speech signals by generalized fractal dimension features," Speech Communication, Vol. 51, issue 12, December 2009, pp. 1206-1223.

[22] W. Wu, T. F. Zheng, M. X. Xu, and H. J. Bao, "Study on speaker verification on emotional speech," INTERSPEECH 2006 – Proceedings of International Conference on Spoken Language Processing (ICSLP), September 2006, pp. 2102-2105.

[23] T. H. Falk and W. Y. Chan, "Modulation spectral features for robust far-field speaker identification," IEEE Transactions on Audio, Speech and Language Processing, Vol. 18, No. 1, January 2010, pp. 90-100.

[24] A. B. Kandali, A. Routray and T. K. Basu, "Emotion recognition from Assamese speeches using MFCC features and GMM classifier," Proc. IEEE Region 10 Conference TENCON 2008, Hyderabad, India, November 2008, pp. 1-5.

[25] X. D. Huang, Y. Ariki, and M. A. Jack, Hidden Markov Models for Speech Recognition, Edinburgh University Press, Great Britain, 1990.

[26] L. R. Rabiner, and B. H. Juang, Fundamentals of Speech Recognition, Prentice Hall, Eaglewood Cliffs, New Jersey, 1983.

[27] D. Ververidis and C. Kotropoulos, "Emotional speech recognition: resources, features, and methods," Speech Communication, Vol. 48, issue 9, September 2006, pp. 1162-1181.

[28] A. Nogueiras, A. Moreno, A. Bonafonte, and J.Mario, "Speech emotion recognition using hidden Markov models," European Conference on Speech Communication and Technology, EUROSPEECH 2001. Aalborg, Denmark, September 2001.

[29] I. Shahin, "Talking condition identification using second-order hidden Markov models," 3$^{rd}$ International Conference on Information & Communication Technologies: from Theory to Applications (ICTTA'08, IEEE Section France), Damascus, Syria, April 2008.

[30] I. R. Murray and J. L. Arnott, "Toward the simulation of emotion in synthetic speech: A review of the literature of human vocal emotion," Journal of the Acoustic Society of America, Vol. 93, No. 2, 1993, pp. 1097-1108.

[31] J.H.L. Hansen and S. Bou-Ghazale, "Getting started with SUSAS: A speech under simulated and actual stress database," EUROSPEECH-97: International Conference on Speech Communication and Technology, Rhodes, Greece, September 1997, pp. 1743-1746.

[32] www.ldc.upenn.edu/Catalog/CatalogEntry.jsp?catalogId=LDC2002S28

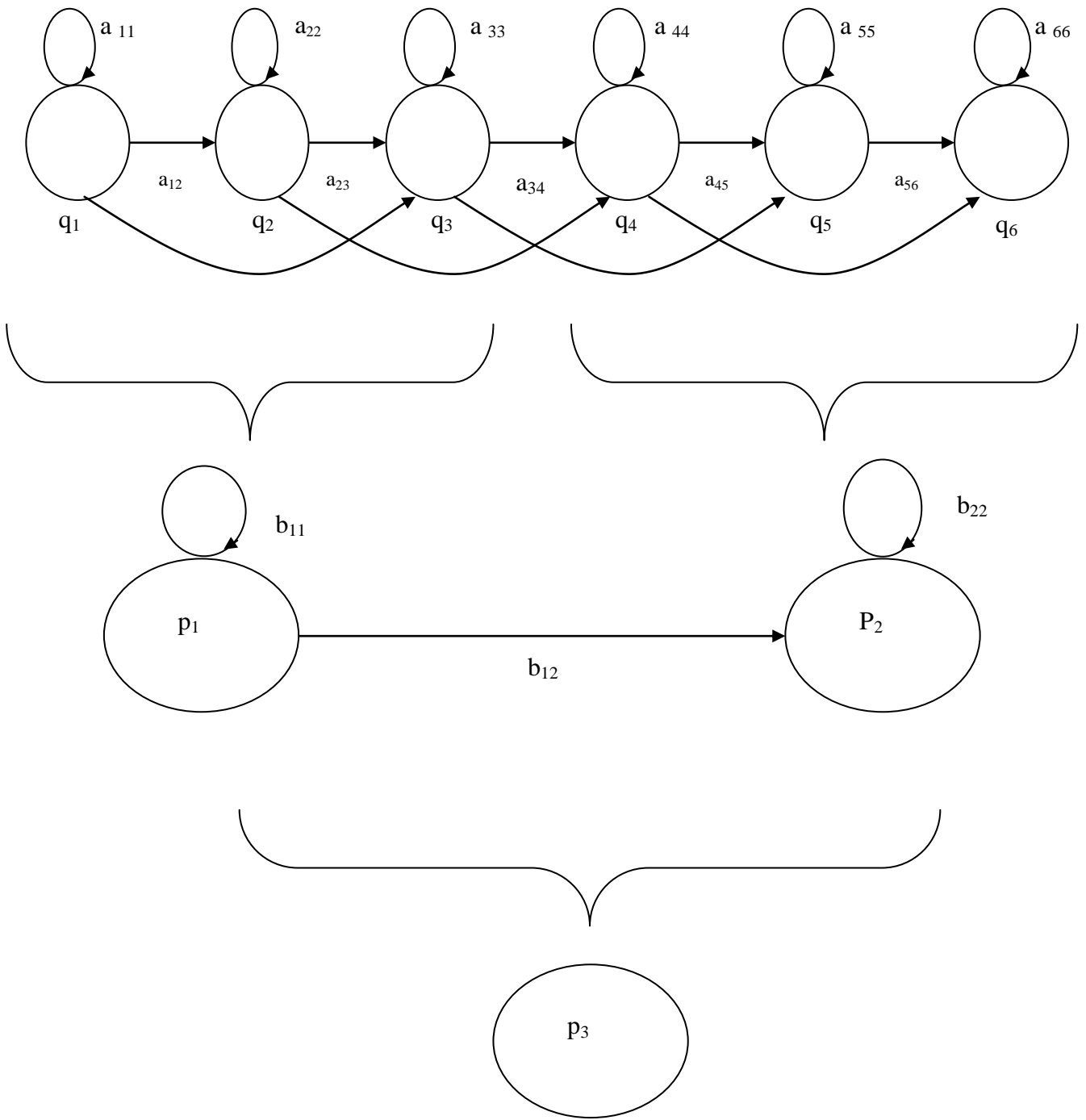

**Fig. 1.** Basic structure of LTRSPHMMs

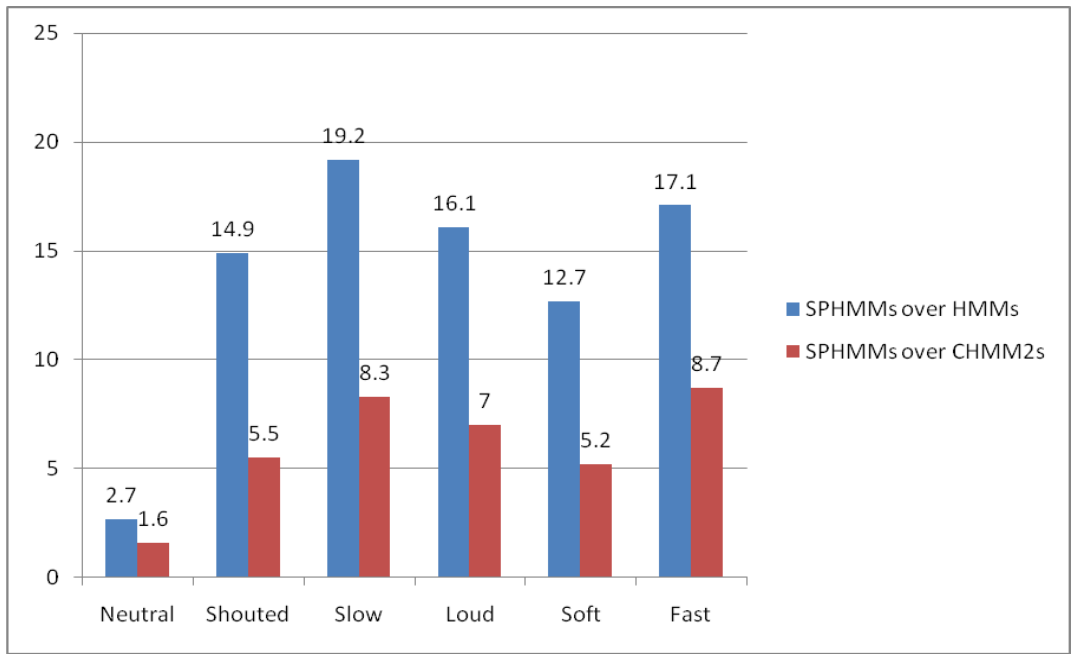

**Fig. 2.** Relative improvement (%) per each stressful talking condition of using SPHMMs ($\alpha = 0.5$) over each of HMMs and CHMM2s

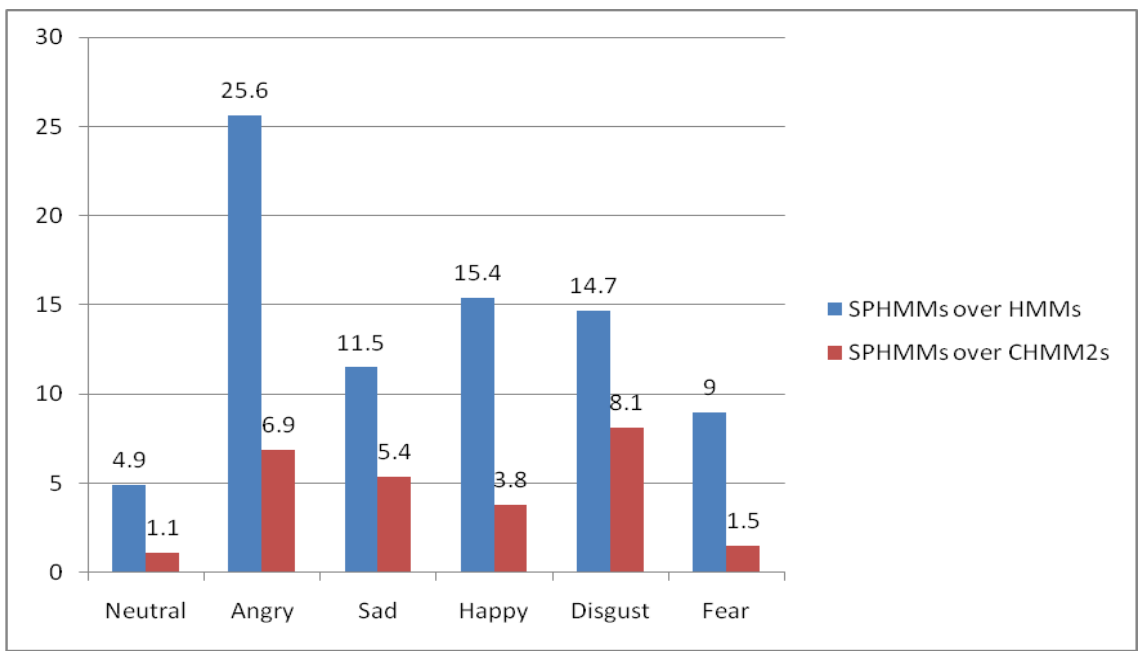

**Fig. 3.** Relative improvement (%) per each emotion of using SPHMMs ($\alpha = 0.5$) over each of HMMs and CHMM2s

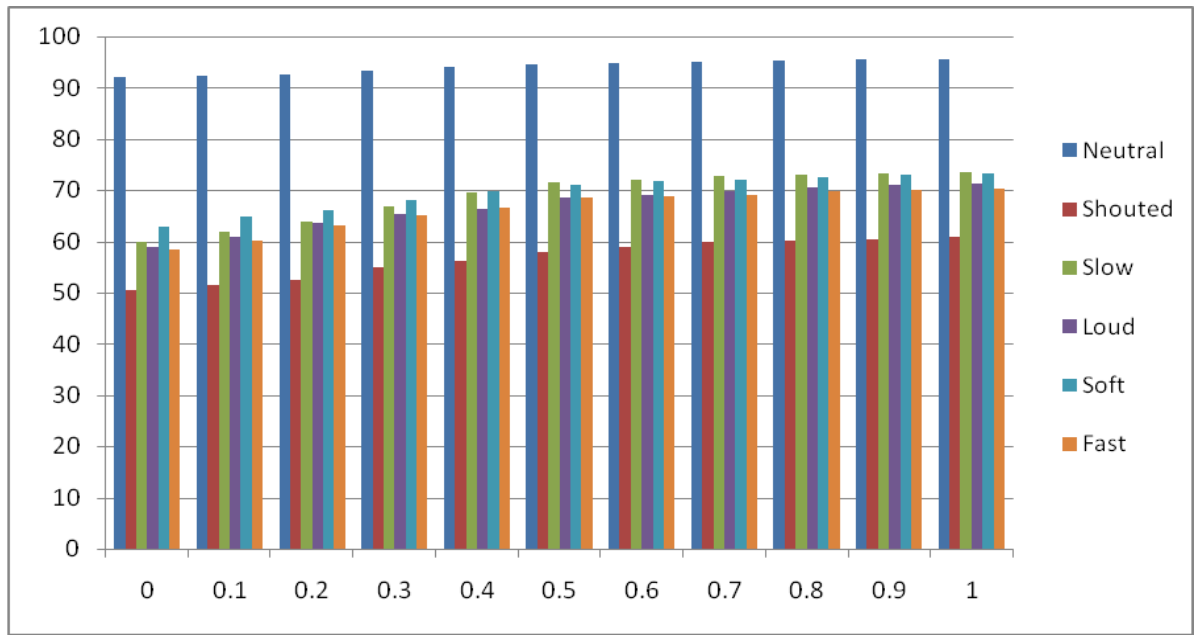

**Fig. 4.** Average stressful talking condition identification performance (%) versus the weighting factor ($\alpha$)

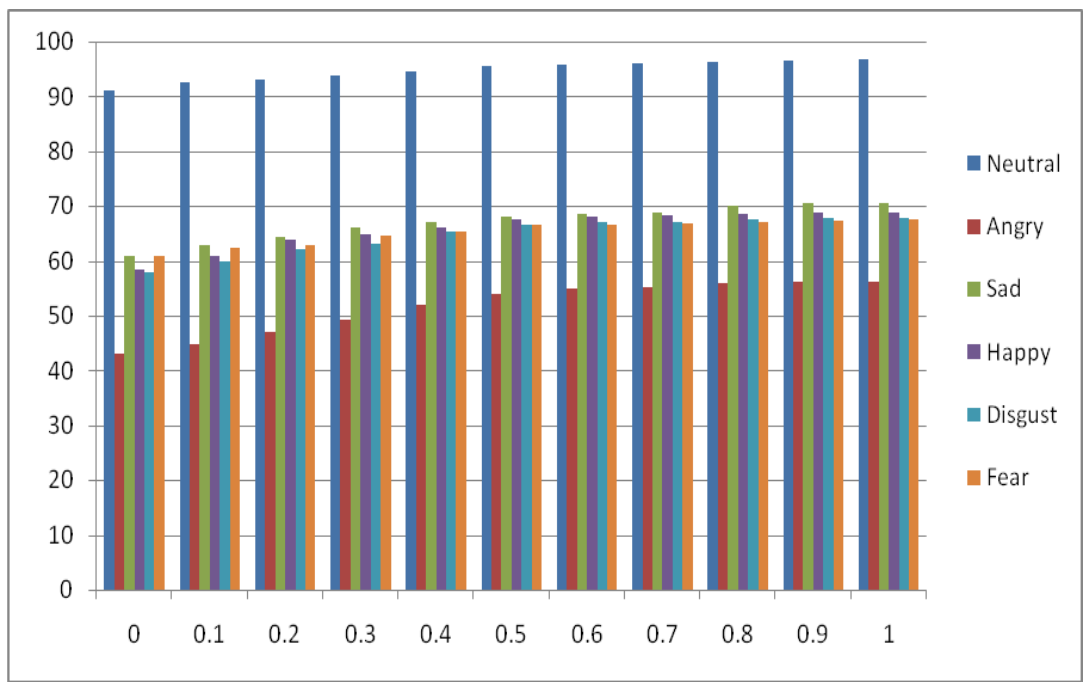

**Fig. 5.** Average emotional talking condition identification performance (%) versus the weighting factor ($\alpha$)

Table 1

Talking condition identification performance in stressful talking environments using the collected database based on HMMs, CHMM2s and SPHMMs ($\alpha = 0.5$)

| Models | Gender | Identification performance under each talking condition (%) | | | | | |
|---|---|---|---|---|---|---|---|
| | | Neutral | Shouted | Slow | Loud | Soft | Fast |
| HMMs | Male | 91 | 51 | 60 | 59 | 64 | 58 |
| | Female | 93 | 50 | 60 | 59 | 62 | 59 |
| | Average | 92 | 50.5 | 60 | 59 | 63 | 58.5 |
| CHMM2s | Male | 93 | 54 | 65 | 65 | 67 | 62 |
| | Female | 93 | 56 | 67 | 63 | 68 | 64 |
| | Average | 93 | 55 | 66 | 64 | 67.5 | 63 |
| SPHMMs | Male | 95 | 58 | 71 | 68 | 70 | 69 |
| | Female | 94 | 58 | 72 | 69 | 72 | 68 |
| | Average | 94.5 | 58 | 71.5 | 68.5 | 71 | 68.5 |

Table 2

Confusion matrix in stressful talking environments using the collected database based on HMMs

|  | Percentage of confusion of a test stressful talking condition with the other stressful talking conditions (%) | | | | | |
|---|---|---|---|---|---|---|
| Model | Neutral | Shouted | Slow | Loud | Soft | Fast |
| Neutral | **92** | 2 | 6 | 3 | 4 | 4 |
| Shouted | 1 | **50.5** | 7 | 20 | 8 | 14.5 |
| Slow | 4 | 6 | **60** | 3.5 | 15 | 4 |
| Loud | 0 | 24.5 | 4 | **59** | 5 | 12 |
| Soft | 3 | 5 | 20 | 6.5 | **63** | 7 |
| Fast | 0 | 12 | 3 | 8 | 5 | **58.5** |

Table 3

Confusion matrix in stressful talking environments using the collected database based on CHMM2s

|  | Percentage of confusion of a test stressful talking condition with the other stressful talking conditions (%) | | | | | |
|---|---|---|---|---|---|---|
| Model | Neutral | Shouted | Slow | Loud | Soft | Fast |
| Neutral | **93** | 4 | 8 | 2 | 5 | 2 |
| Shouted | 1 | **55** | 3 | 17 | 5 | 14 |
| Slow | 3 | 5 | **66** | 3 | 12 | 5 |
| Loud | 1 | 21 | 6 | **64** | 6 | 11 |
| Soft | 1 | 5 | 15 | 6 | **67.5** | 5 |
| Fast | 1 | 10 | 2 | 8 | 4.5 | **63** |

Table 4

Confusion matrix in stressful talking environments using the collected database based on SPHMMs ($\alpha = 0.5$)

|  | Percentage of confusion of a test stressful talking condition with the other stressful talking conditions (%) | | | | | |
|---|---|---|---|---|---|---|
| Model | Neutral | Shouted | Slow | Loud | Soft | Fast |
| Neutral | **94.5** | 2 | 5 | 2 | 6 | 2 |
| Shouted | 1 | **58** | 3 | 16 | 2 | 11 |
| Slow | 4.5 | 6 | **71.5** | 3.5 | 14 | 5 |
| Loud | 0 | 20 | 2 | **68.5** | 3 | 10 |
| Soft | 0 | 5 | 15 | 6 | **71** | 3.5 |
| Fast | 0 | 9 | 3.5 | 4 | 4 | **68.5** |

Table 5

Emotion identification performance in emotional talking environments using the collected database based on HMMs, CHMM2s and SPHMMs ($\alpha = 0.5$)

| Models | Gender | Identification performance under each emotion (%) | | | | | |
|---|---|---|---|---|---|---|---|
| | | Neutral | Angry | Sad | Happy | Disgust | Fear |
| HMMs | Male | 90 | 42 | 60 | 58 | 58 | 62 |
| | Female | 92 | 44 | 62 | 59 | 58 | 60 |
| | Average | 91 | 43 | 61 | 58.5 | 58 | 61 |
| CHMM2s | Male | 95 | 50 | 64 | 64 | 62 | 65 |
| | Female | 94 | 51 | 65 | 66 | 61 | 66 |
| | Average | 94.5 | 50.5 | 64.5 | 65 | 61.5 | 65.5 |
| SPHMMs | Male | 96 | 54 | 67 | 67 | 66 | 66 |
| | Female | 95 | 54 | 69 | 68 | 67 | 67 |
| | Average | 95.5 | 54 | 68 | 67.5 | 66.5 | 66.5 |

Table 6

Confusion matrix in emotional talking environments using the collected database based on HMMs

| | Percentage of confusion of a test emotion with the other emotions (%) | | | | | |
|---|---|---|---|---|---|---|
| Model | Neutral | Angry | Sad | Happy | Disgust | Fear |
| Neutral | **91** | 2 | 3 | 1.5 | 2 | 4 |
| Angry | 1 | **43** | 15 | 11 | 20 | 11 |
| Sad | 3 | 14 | **61** | 7 | 6 | 12 |
| Happy | 2 | 6 | 4 | **58.5** | 4 | 3 |
| Disgust | 1 | 23 | 8 | 10 | **58** | 9 |
| Fear | 2 | 12 | 9 | 12 | 10 | **61** |

Table 7

Confusion matrix in emotional talking environments using the collected database based on CHMM2s

| | Percentage of confusion of a test emotion with the other emotions (%) | | | | | |
|---|---|---|---|---|---|---|
| Model | Neutral | Angry | Sad | Happy | Disgust | Fear |
| Neutral | **94.5** | 2.5 | 1 | 3 | 2 | 2 |
| Angry | 0 | **50.5** | 13.5 | 9 | 15.5 | 10 |
| Sad | 2.5 | 11 | **64.5** | 7 | 8 | 12 |
| Happy | 1 | 3 | 2 | **65** | 2 | 1 |
| Disgust | 1 | 23 | 9 | 6 | **61.5** | 9.5 |
| Fear | 1 | 10 | 10 | 10 | 11 | **65.5** |

Table 8

Confusion matrix in emotional talking environments using the collected database based on SPHMMs ($\alpha = 0.5$)

| | Percentage of confusion of a test emotion with the other emotions (%) | | | | | |
|---|---|---|---|---|---|---|
| Model | Neutral | Angry | Sad | Happy | Disgust | Fear |
| Neutral | **95.5** | 1 | 3 | 7 | 2 | 3.5 |
| Angry | 0 | **54** | 10 | 5 | 15.5 | 9 |
| Sad | 1.5 | 10 | **68** | 4.5 | 6 | 10 |
| Happy | 2 | 5 | 3 | **67.5** | 4 | 2 |
| Disgust | 0 | 20 | 5.5 | 8 | **66.5** | 9 |
| Fear | 1 | 10 | 10.5 | 8 | 6 | **66.5** |

Table 9

Talking condition identification performance in stressful talking environments

using SUSAS database based on HMMs, CHMM2s and SPHMMs ($\alpha = 0.5$)

| Models | Gender | Identification performance under each talking condition (%) | | | | | |
|---|---|---|---|---|---|---|---|
| | | Neutral | Angry | Slow | Loud | Soft | Fast |
| HMMs | Male | 93 | 53 | 61 | 59 | 64 | 60 |
| | Female | 93 | 51 | 61 | 60 | 64 | 59 |
| | Average | 93 | 52 | 61 | 59.5 | 64 | 59.5 |
| CHMM2s | Male | 93 | 56 | 66 | 65 | 69 | 63 |
| | Female | 94 | 57 | 68 | 65 | 68 | 64 |
| | Average | 93.5 | 56.5 | 67 | 65 | 68.5 | 63.5 |
| SPHMMs | Male | 95 | 59 | 73 | 69 | 72 | 69 |
| | Female | 95 | 58 | 72 | 69 | 73 | 69 |
| | Average | 95 | 58.5 | 72.5 | 69 | 72.5 | 69 |

Table 10

Emotion identification performance in emotional talking environments using Emotional Prosody database based on HMMs, CHMM2s and SPHMMs ($\alpha = 0.5$)

| Models | Gender | Identification performance under each emotion (%) | | | | | |
|---|---|---|---|---|---|---|---|
| | | Neutral | Hot Anger | Sadness | Happiness | Disgust | Panic |
| HMMs | Male | 92 | 47 | 63 | 59 | 60 | 62 |
| | Female | 92 | 45 | 62 | 59 | 60 | 62 |
| | Average | 92 | 46 | 62.5 | 59 | 60 | 62 |
| CHMM2s | Male | 95 | 53 | 64 | 66 | 62 | 67 |
| | Female | 95 | 51 | 67 | 66 | 63 | 66 |
| | Average | 95 | 52 | 65.5 | 66 | 62.5 | 66.5 |
| SPHMMs | Male | 96 | 56 | 68 | 69 | 67 | 66 |
| | Female | 95 | 56 | 69 | 68 | 69 | 66 |
| | Average | 95.5 | 56 | 68.5 | 68.5 | 68 | 66 |